# Tuning light-matter interaction of near-infrared nanoplasmonic scintillators

**Michał Makowski[1,2,*], Dominik Kowal[1], and Muhammad Danang Birowosuto[1]**

[1]Łukasiewicz Research Network - PORT Polish Center for Technology Development, Wrocław, 54-066, Poland
[2]Institute of Physics, Faculty of Physics, Astronomy and Informatics, Nicolaus Copernicus University, Grudziadzka 5, 87-100 Torun, Poland
[*]michal.makowski@port.lukasiewicz.gov.pl

## ABSTRACT

Nanoplasmonic modification of scintillation has so far been explored mainly in the weak-coupling regime, where changes in the local density of optical states enhance radiative recombination via Purcell-type rate engineering. By contrast, strong light-matter coupling generates hybrid states that modify emission dynamics beyond simple decay-rate acceleration, but its implications for scintillator nanocrystals (NCs) under ionizing radiation remain poorly understood. All of these effects are beneficial for near-infrared scintillators, which are typically slow and have low brightness. Here, we present a quantum-optical framework to investigate how near-infrared scintillator NCs coupled to nanoplasmonic antennas evolve from weak coupling toward strong light-matter coupling. We compare broad- and narrow-antenna platforms with single and periodic Au nanorods and benchmark them against conductive plasmonic antennas based on indium tin oxide and graphene. As representative emitters, we consider wide-band PbS NCs and narrow-band cubic $Lu_2O_3:Er^{3+}$ scintillators. The calculations show that the onset of strong-coupling signatures is jointly governed by emitter dephasing and the antenna linewidth, with narrow-band emitters coupled to spectrally narrow antennas providing the most favorable conditions. Among the platforms considered, graphene gives the lowest threshold ($g$ = 4 meV) for observable coherent exchange owing to its ultranarrow antenna linewidth ($\kappa$ = 3.5 meV). These results identify near-infrared conductive nanoantennas, particularly graphene-based ones, as promising platforms for accessing hybrid scintillation regimes relevant to radiation detection.

## Introduction

Nanoplasmonic antennas and metasurface architectures allow control over light-matter interactions at the nanoscale. Such control is increasingly being explored in luminescent materials, where the local photonic environment can modify both emission efficiency and decay dynamics. This is particularly relevant for scintillators, which are key components of radiation detection systems and whose light yield and temporal response directly determine energy and timing resolution[1,2]. In conventional scintillators, the emission kinetics are governed by intrinsic radiative and nonradiative recombination rates and are therefore only weakly tunable through material composition alone[3]. As a result, accelerating scintillation without sacrificing light yield remains a long-standing challenge.

Recent work in nanoplasmonics has shown that spontaneous emission can be modified by tailoring the local density of optical states (LDOS), giving rise to so-called rate engineering in the weak-coupling regime[4–6]. In scintillators, plasmonic and dielectric nanostructures have already been used to enhance radiative recombination through the Purcell effect, leading to faster emission under both optical and ionizing excitation[7–9]. In this regime, light and matter remain distinct, and the nanoplasmonic structure modifies the emission rate rather than creating new hybrid states.

At the same time, alternative plasmonic materials have attracted increasing attention, as they expand the accessible design space for nanoplasmonic antennas, particularly in spectral regions where gold and silver are less effective[10–13]. In particular, conductive platforms such as transparent conducting oxides and graphene are especially promising in the near-infrared, where they can support tunable optical modes with reduced losses compared to conventional noble-metal plasmonics. Moreover, both emission acceleration and light-yield enhancement are critical for near-infrared scintillators[14]. The near-infrared thus constitutes a particularly relevant spectral window for exploring nanoplasmonic control of scintillation. From a scintillator design perspective, this region is also attractive because lower-band-gap materials can, in principle, generate a larger number of electron–hole pairs per unit of deposited energy, which is favorable for achieving high light yield, although the overall scintillation efficiency is not determined solely by the band gap. Furthermore, if the emission spectrum is shifted into the infrared, especially toward wavelengths where silicon absorption is reduced, photons emitted away from the scintillating layer could, in principle, be collected remotely, potentially even through additional semiconductor-based tracking layers. Such a configuration may open new opportunities in detector design and functionality. Taken together, these considerations make near-infrared scintillator nanocrystals (NCs) a compelling platform for investigating whether nanoplasmonic coupling can extend scintillation beyond rate engineering into regimes of coherent light–matter hybridization.

Beyond rate engineering lies the strong-coupling regime, where coherent energy exchange between an emitter and an optical mode gives rise to hybrid light-matter states separated by the vacuum Rabi splitting[15–17]. This regime has been widely studied in semiconductor nanostructures, nanoplasmonic antennas, and related nanophotonic platforms, most often under optical excitation[18–20]. By contrast, its role in scintillation under ionizing radiation remains much less clear. Despite recent demonstrations of strong light-matter coupling in bulk nanoplasmonic perovskite scintillators[21], a unified theoretical description linking this regime to scintillation dynamics remains lacking.

Ionizing excitation differs fundamentally from optical pumping. Under ionizing radiation, carrier populations are generated incoherently and far from thermal equilibrium, and the observed emission reflects the interplay of carrier relaxation, recombination, and coupling to plasmonic or photonic modes[1,3]. For this reason, signatures of strong coupling in radioluminescence cannot be inferred directly from conventional optical measurements, and a unified framework that connects the weak- and strong-coupling regimes in scintillators remains missing.

Here, we develop a theoretical description of the transition from Purcell-enhanced near-infrared scintillation in the weak-coupling regime to the onset of strong light-matter coupling under ionizing excitation. Using an open quantum system approach[22,23], we show how increasing the emitter-mode coupling strength drives a continuous crossover from rate-modified emission to the formation of hybrid states, and we identify its signatures in the temporal and spectral response of scintillators. Our results provide design rules for accessing strong coupling in scintillator NCs and clarify how this regime modifies scintillation dynamics.

## Results

### Theoretical model

The emitter-antenna system was described within the framework of open quantum systems using a driven-dissipative Jaynes-Cummings model[15,22,23]. The emitter was treated as an effective two-level system coupled to a single confined optical mode representing the nanoplasmonic antenna. In the rotating frame used in the simulations, the Hamiltonian takes the form

$$\hat{H} = \hbar\Delta_c \hat{a}^\dagger \hat{a} + \hbar g \left(\hat{a}^\dagger \hat{\sigma} + \hat{a}\hat{\sigma}^\dagger\right), \tag{1}$$

where $\hat{a}$ ($\hat{a}^\dagger$) are the annihilation (creation) operators of the optical mode, $\hat{\sigma}$ ($\hat{\sigma}^\dagger$) are the lowering (raising) operators of the emitter, $g$ is the light-matter coupling strength, and $\Delta_c = \omega_c - \omega_e$ is the detuning between the antenna $\omega_c$ and emitter transition energies $\omega_e$. Dissipative dynamics were included through a Lindblad master equation,

$$\frac{d\hat{\rho}}{dt} = -\frac{i}{\hbar}[\hat{H},\hat{\rho}] + \sum_i \mathscr{L}_i[\hat{\rho}], \tag{2}$$

where the Lindblad terms account for antenna losses, emitter decay, pure dephasing, and incoherent pumping. The transition from weak to strong coupling is governed by the competition between coherent exchange, quantified by $g$, and the relevant dissipative rates[18]. In the weak-coupling regime, the interaction mainly modifies the emission dynamics, whereas at larger coupling strengths coherent exchange gives rise to oscillatory temporal behavior (Rabi oscillations) and the emergence of a split emission spectrum.

The temporal response was analyzed through the first-order correlation function of the detected field, while the spectral response was obtained from steady-state two-time correlation functions of the antenna and emitter channels and constructed as an incoherent sum of their respective contributions. Further details of the collapse operators, detected-field definition, spectral decomposition, and numerical implementation are provided in the Supplementary Information in Quantum-optical model and numerical implementation section.

### Simulation Results

To identify the conditions under which scintillator NCs evolve from Purcell-enhanced emission to strong light-matter coupling, we compare model systems that differ in two key parameters: the spectral linewidth of the emitter and the linewidth of the antenna. Two representative emitters were therefore selected: PbS NCs as an example of a wide-band scintillator (WBS) with emitter transition energy $\omega_e = 1.319$ eV and pure dephasing rate $\gamma_\phi = 75$ meV[24], and the $Er^{3+}$ $^4I_{13/2} \rightarrow {}^4I_{15/2}$ transition in cubic $Lu_2O_3$:$Er^{3+}$ NCs as an example of a narrow-band scintillator (NBS) with $\omega_e = 0.809$ eV, $\gamma_\phi = 10$ meV[25]. For simplicity, we determined the quantum efficiencies of both emitters are unity. These emitters were combined with antennas supporting either broad or narrow optical modes, allowing us to isolate the joint effect of emitter and antenna linewidth on the temporal and spectral signatures of the coupling regime.

Fig. 1 summarizes the concept of this work. Bare NCs under X/$\gamma$-ray excitation exhibit a conventional radiative response with a single emission band (Fig. 1a). When coupled to a broad optical mode, represented here by a single Au nanorod (Fig. 1b), the local photonic environment modifies the emission dynamics through the Purcell effect, leading to a faster temporal response and enhanced emission intensity while the system remains in the weak-coupling regime and no hybrid light-matter states are formed[18,26]. In a periodic nanorod array, the plasmonic resonance narrows due to collective lattice effects, which reduce optical losses and make strong-coupling signatures spectrally and temporally resolvable[27,28]. In this case, the temporal response develops Rabi oscillations, and the detected spectrum evolves into a split polaritonic doublet (Fig. 1c)[18,27,29].

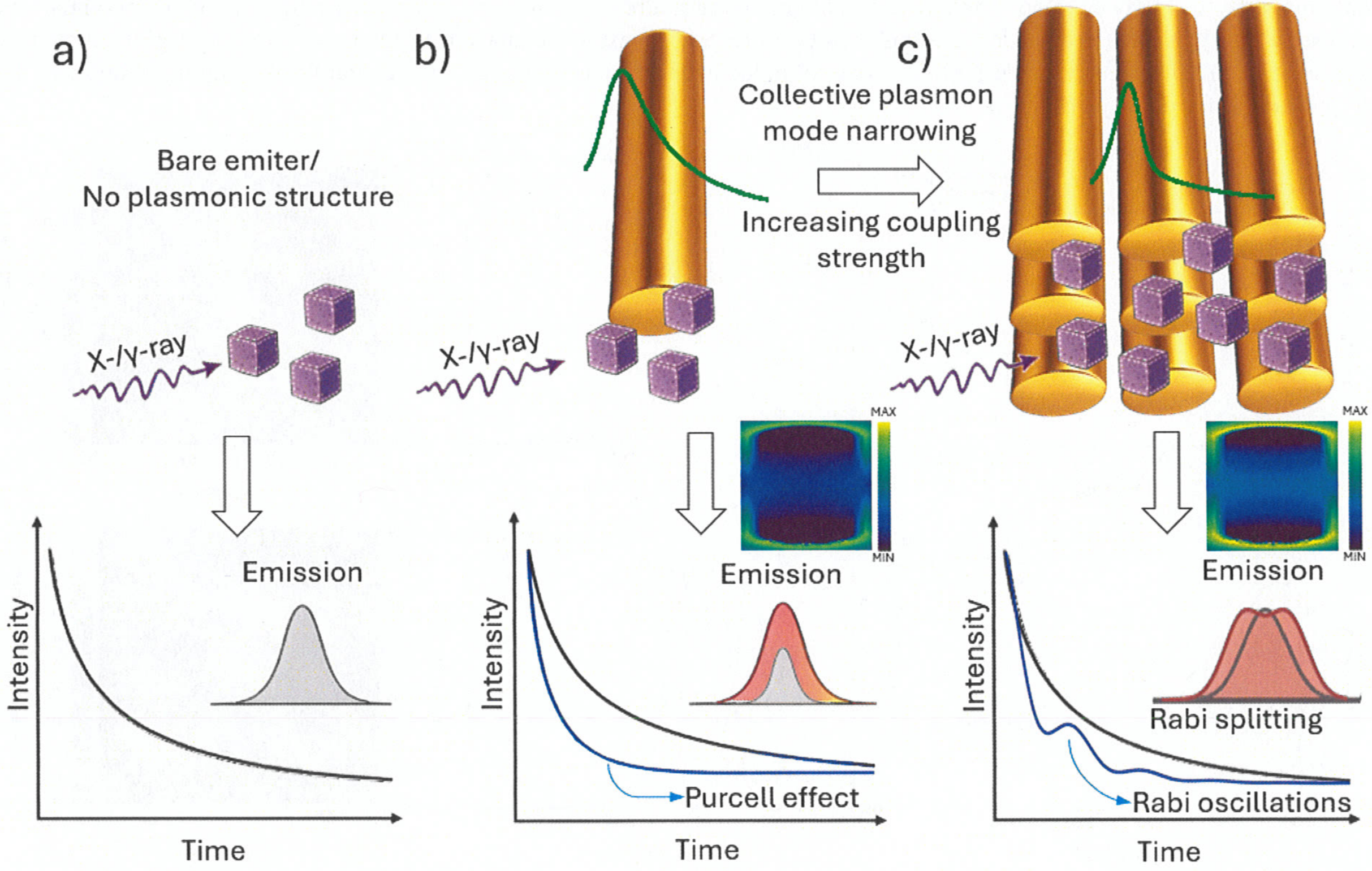

**Figure 1.** Conceptual illustration of scintillation coupled to plasmonic antennas or nanostructures. **a)** Bare nanocrystals (NCs) excited by X/$\gamma$-ray show conventional radiative decay and a single emission band. **b)** NCs coupled to the plasmonic mode of a single Au nanorod (green curve) result in Purcell-enhanced emission, observed as increased intensity and faster decay. **c** NCs coupled to the periodic array of nanorods, collective plasmon modes become narrower, increasing the coupling strength to the strong coupling regime, verified by emission Rabi splitting and damped Rabi oscillations in the decay dynamics. Colorful insets in (b) and (c) present cross-section mode profiles of a single Au nanorod (b) and a nanorod Array (c).

To translate the concept picture of Fig. 1 into quantitative design rules, we next compare the selected emitters with nanoplasmonic antennas of different linewidths. As a broadband antenna, we consider a single Au nanorod spectrally matched to the emitter transition. Gold nanorods were chosen as a reference plasmonic platform because they are among the most widely used and best understood anisotropic plasmonic antennas. For the WBS case, we use a nanorod with dimensions $L \times D = 130 \times 30$ nm, whereas for the NBS case, we use a nanorod with dimensions $240 \times 30$ nm (see Supplementary Fig. S1). As a narrowband antenna, we consider periodic 2D arrays of the same nanorods, which support collective plasmonic modes with reduced linewidths compared to the corresponding single-particle resonances. The array periods were set to 600 nm $\times$ 880 nm (X $\times$ Y) for the 130 $\times$ 30 nm nanorods and 980 nm $\times$ 1300 nm (X $\times$ Y) for the 240 $\times$ 30 nm nanorods. The corresponding resonance energies $\omega_c$, linewidths, and detunings extracted from the simulated scattering spectra (Supplementary Fig. S2 and Fig. S3) are summarized in Supplementary Table S1.

Simulations of the WBS coupled to Au nanorods show that the visibility of strong-coupling signatures depends strongly on the antenna linewidth. For the wider antenna (Fig. 2 a,c), the spectrum at $g = 40$ meV does not even begin to develop a split profile. In contrast, for the narrower band of the antenna (Fig. 2 b,d), a distinct spectral splitting is already visible at the same coupling strength. This indicates that a narrower band of the antenna is more favorable for resolving strong-coupling features in a broadband emitter. With a further increase in $g$, the splitting becomes more pronounced in both cases, but it remains consistently better resolved in the narrower band of the antenna. The same trend is observed in the temporal response. Clear Rabi oscillations appear only at the highest coupling strength, $g = 140$ meV, for both antennas. However, they are more pronounced for the narrower band of the antenna, which shows a deeper first minimum in $g^{(1)}(\tau)$. Overall, these results show that reducing the antenna linewidth improves both the spectral and temporal visibility of coherent light-matter coupling. This behavior is also evident in the detuning-dependent spectral maps shown in Fig. 2e,f. For the broadband antenna, the emission remains dominated by a single broadened branch with only a weak indication of mode repulsion, whereas for the narrowband antenna, a more clearly developed anticrossing pattern emerges already at the same coupling strength. A broader comparison across selected $g$ values is provided in Supplementary Fig. S4. An extended analysis of the weak-coupling regime, including Purcell-type enhancement trends prior to the onset of mode hybridization, is provided in Section Weak-coupling analysis of the Supporting Information and Fig. S5.

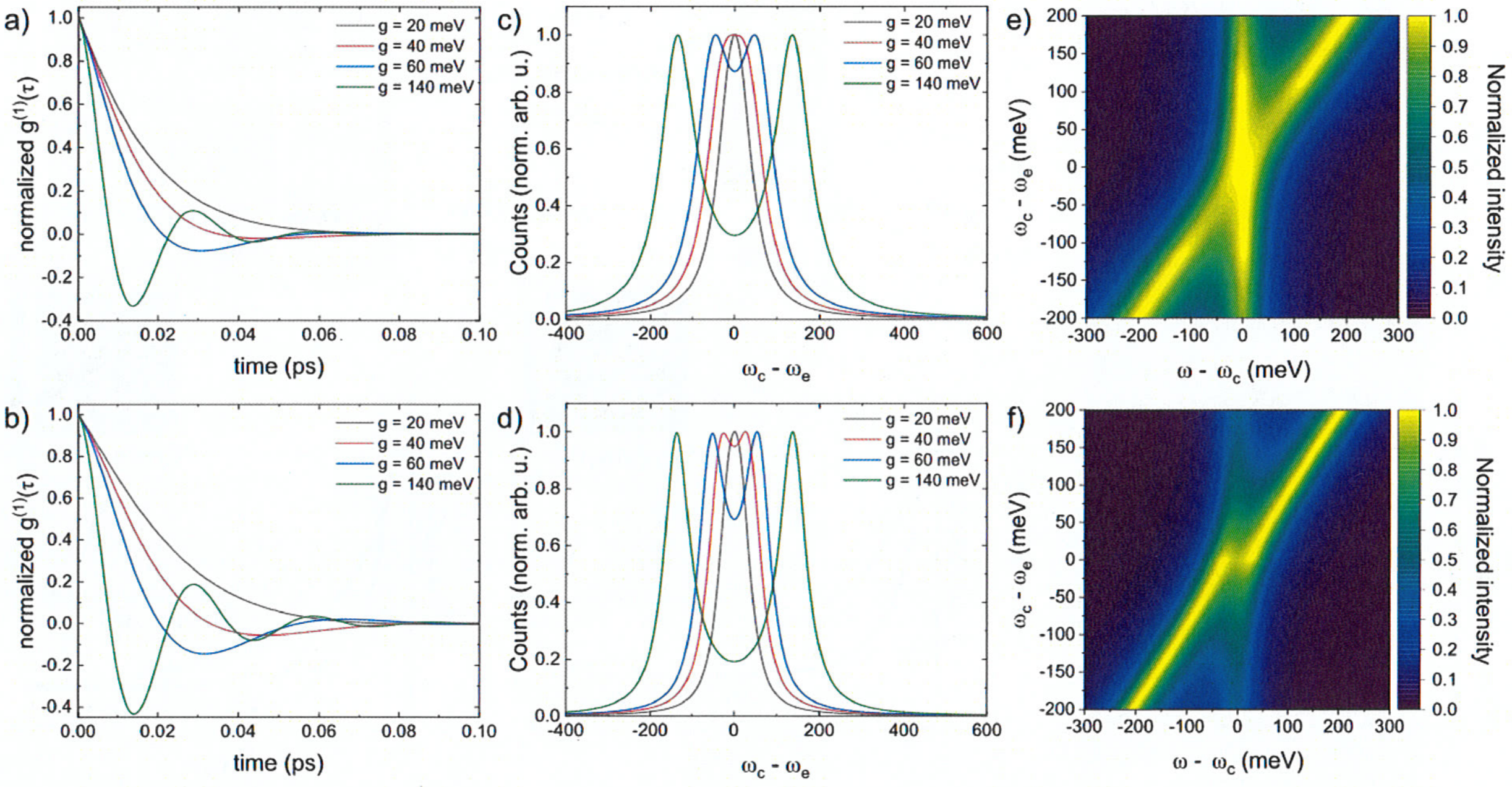

**Figure 2.** Simulated temporal and spectral response of a wide-band scintillator (WBS) coupled to Au nanoantennas with different optical linewidths. **a,b**, Normalized first-order coherence function $g^{(1)}(\tau)$ for selected coupling strengths $g$. **c,d**, Corresponding normalized detected spectra $S_{\text{det}}(\omega)$. **e,f**, Detuning-dependent spectral maps of the detected emission for $g = 40$ meV. **a,c,e**, Broad Au antenna mode. **b,d,f**, Narrow Au antenna mode.

The simulations performed for the NBS coupled to Au nanorods reveal that the antenna linewidth plays an even more decisive role than in the WBS case. For the wider antenna band (Fig. 3 a,c), the spectrum at $g = 30$ meV does not yet show a clearly resolved doublet, but only a weak shoulder-like deformation of the emission band. By contrast, for the narrower antenna band (Fig. 3 b,d), the same coupling strength already produces a distinct and well-resolved splitting. This demonstrates that linewidth narrowing strongly promotes the spectral visibility of strong-coupling signatures for a narrow-band emitter. As $g$ increases, the split peaks become more pronounced in both cases, yet they remain consistently sharper and better separated in the narrower band case. A similar difference is observed in the temporal domain, where Rabi oscillations are clearly more pronounced in the narrower band of the antenna, with more visible oscillation cycles before decay. Rabi oscillations are observed at lower coupling strength for the narrower band of the antenna, appearing already at $g = 80$ meV, whereas the broader band of the antenna requires $g = 100$ meV. Compared with the WBS system, the NBS case therefore exhibits a more explicit transition into the strong-coupling regime, particularly when paired with a narrowband antenna. This contrast is also reflected in the detuning-dependent spectral maps in Fig. 3e,f. While the broadband antenna shows only a weakly developed anticrossing, the narrowband antenna exhibits a much clearer separation of the hybrid branches near zero detuning. The same trend over a wider range of coupling strengths is presented in Supplementary Fig. S6.

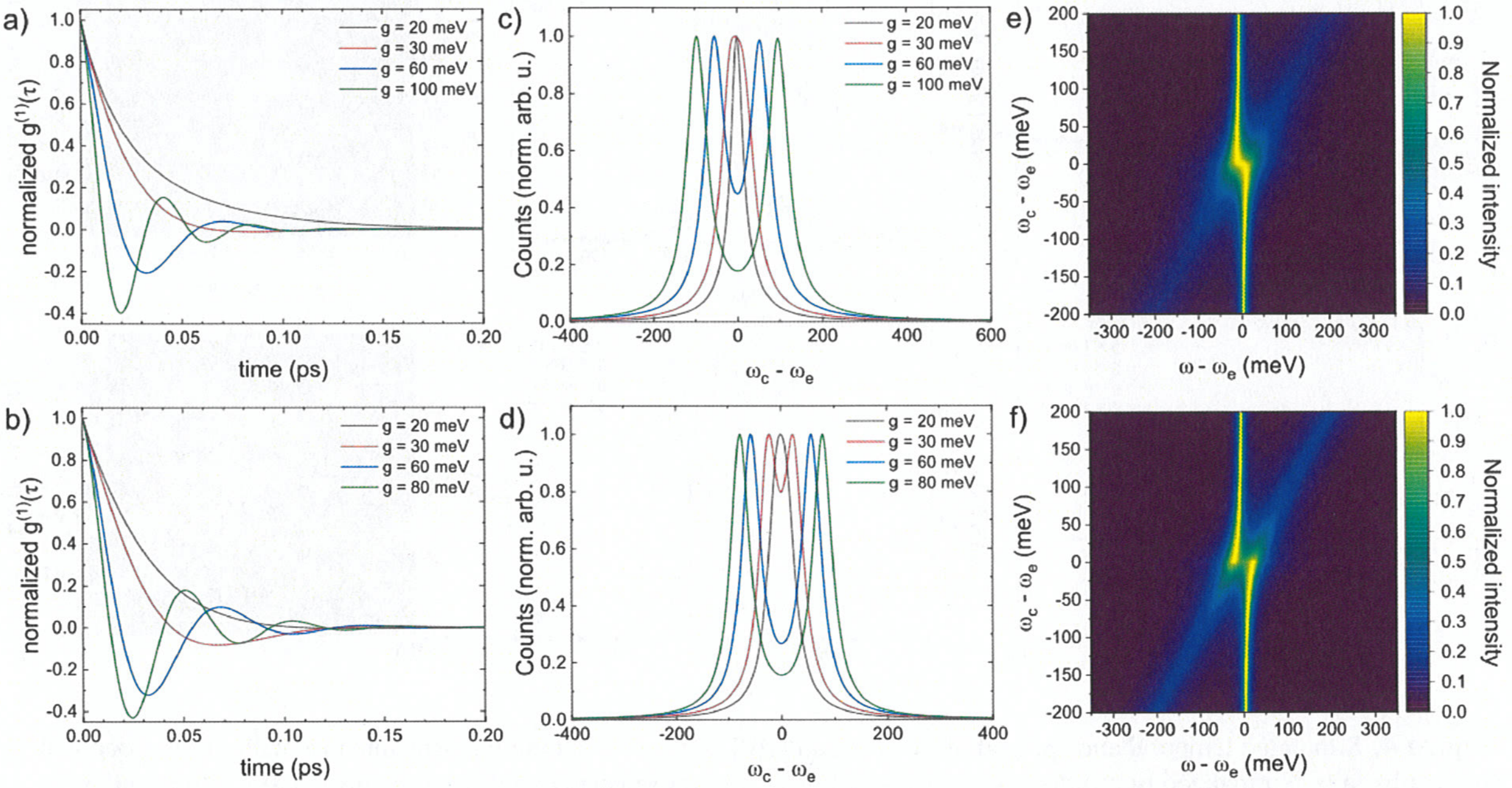

**Figure 3.** Simulated temporal and spectral response of an narrow-band scintillator (NBS) coupled to Au nanoantennas with different optical linewidths. **a,b**, Normalized first-order coherence function $g^{(1)}(\tau)$ for selected coupling strengths $g$. **c,d**, Corresponding normalized detected spectra $S_{\rm det}(\omega)$. **a,c** Broad Au antenna mode. **b,d** Narrow Au antenna mode. **e,f**, Detuning-dependent spectral maps of the detected emission for $g = 30$ meV. **a,c,e**, Broad Au antenna mode. **b,d,f**, Narrow Au antenna mode.

The results discussed above show that the most favorable conditions for the emergence of strong-coupling signatures are obtained for the combination of an NBS and a spectrally narrow antenna mode, for which the transition from Purcell-enhanced emission to the strong-coupling regime occurs already at the smallest values of $g$. Motivated by these findings, we next examine alternative conductive antenna platforms based on indium tin oxide (ITO) and graphene. In recent years, transparent conducting oxides have attracted considerable attention as emerging plasmonic materials, particularly in the near-infrared, where ITO supports tunable plasmonic and epsilon-near-zero optical responses and offers a promising route beyond conventional noble-metal nanostructures[13,30–33]. In parallel, graphene provides an even more strongly confined and spectrally narrower optical mode[12,34–37]. Accordingly, we performed analogous simulations for the NBS coupled to a spherical ITO antenna ($r = 45$ nm) and to a graphene flake with a lateral size of $13 \times 13$ nm and thickness of 1 nm (Supplementary Fig. S7). The scattering spectra of such antennas are presented in Supplementary Fig. S8, with the extracted data shown in Tab. S2. The corresponding spectral maps and temporal response are presented in Fig. 4.

For the ITO antenna, $g = 40$ meV is enough to observe initial spectral splitting, while $g = 140$ meV is required to see the Rabi oscillations. An even more pronounced effect is obtained for the graphene antenna. Owing to the extreme antenna band narrowing ($\kappa = 3.5$ meV, Supplementary Fig. S8), the NBS-graphene system enters the strong-coupling regime already at $g = 4$ meV, far below the coupling strengths required for the metallic nanoantennas considered above. This ultranarrow mode not only produces an early and clearly resolved spectral splitting, but also sustains damped oscillations in $g^{(1)}(\tau)$ out to nearly 2 ps, indicating a substantially prolonged coherent light-matter energy exchange. The same trend is reflected in the detuning-dependent spectral maps shown in Fig. 4e,f. While the ITO antenna band already exhibits a localized anticrossing near zero detuning, the graphene antenna produces a much sharper, more clearly separated hybrid-mode pattern, consistent with its much lower coupling threshold. A broader comparison of the detuning-dependent response for both alternative antennas is provided in Supplementary Fig. S9.

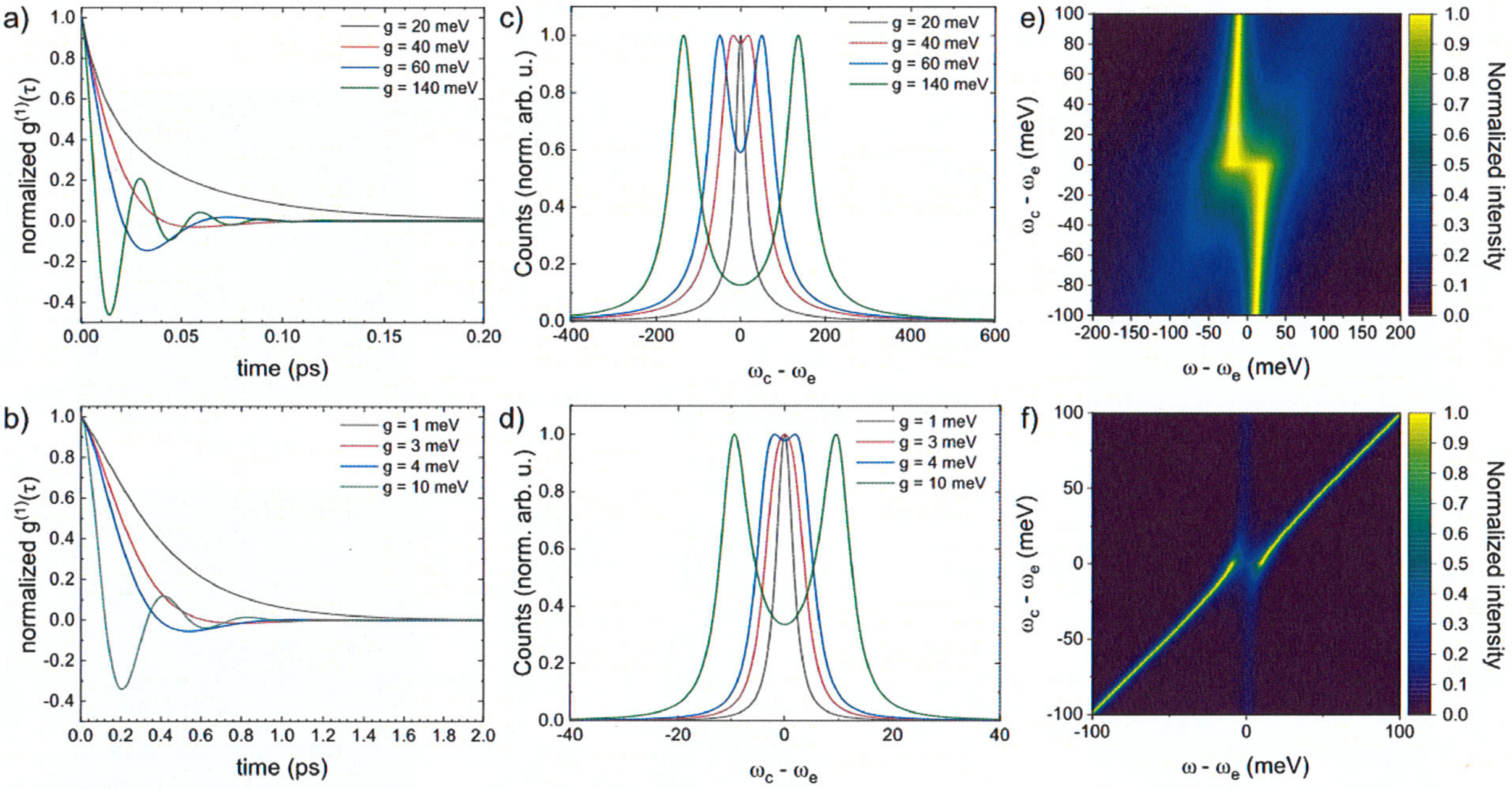

**Figure 4.** Simulated temporal and spectral response of an NBS coupled to alternative nanoantennas with different optical linewidths. **a,b**, Normalized first-order coherence function $g^{(1)}(\tau)$ for selected coupling strengths $g$. **c,d**, Corresponding normalized detected spectra $S_{\mathrm{det}}(\omega)$. **e,f**, Detuning-dependent spectral maps of the detected emission for $g$ = 40 meV (e) and $g$ = 10 meV (f). **a,c,e**, ITO antenna mode. **b,d,f**, graphene antenna mode.

Within the present model, the extracted enhancement remains broadly comparable for the two emitters, since the effective emitter linewidth is dominated by pure dephasing, whereas the radiative channel contributes to the detected field only through its square-root amplitude. Consequently, differences in the bare emitter decay rate translate only weakly into the overall enhancement trends, particularly for the small variation in cavity linewidth considered here. The largest Purcell factor ($F_P$) is obtained for cavities that remain outside the strong-coupling regime up to the highest coupling strengths, reaching $F_P$ = 850 ± 10. Qualitatively, this enhancement is expected to affect the emitters somewhat differently: in lower-QE systems (WBS), it should contribute more strongly to emission efficiency, whereas in higher-QE (NBS) long-lifetime emitters, it should be expressed more directly through lifetime shortening.

Figure 5 summarizes the parameter space explored in this work in terms of emitter dephasing, $\gamma_\phi$, and antenna linewidth, $\kappa$. The colored background indicates the strong-coupling threshold, $g_{\mathrm{th}}$, required for the onset of clearly resolved strong-coupling signatures. Black markers denote the Au-based reference systems discussed above, including single nanorods and nanorod arrays, whereas the red markers highlight the alternative conductive antennas, namely ITO and graphene. The annotations next to each point summarize the simulated coupling strengths and indicate whether strong-coupling features are absent or resolved for the selected $g$ values. This overview highlights that the most favorable conditions are obtained for narrow-band emitters combined with ultranarrow antenna modes, with the graphene antenna representing the most extreme case considered here.

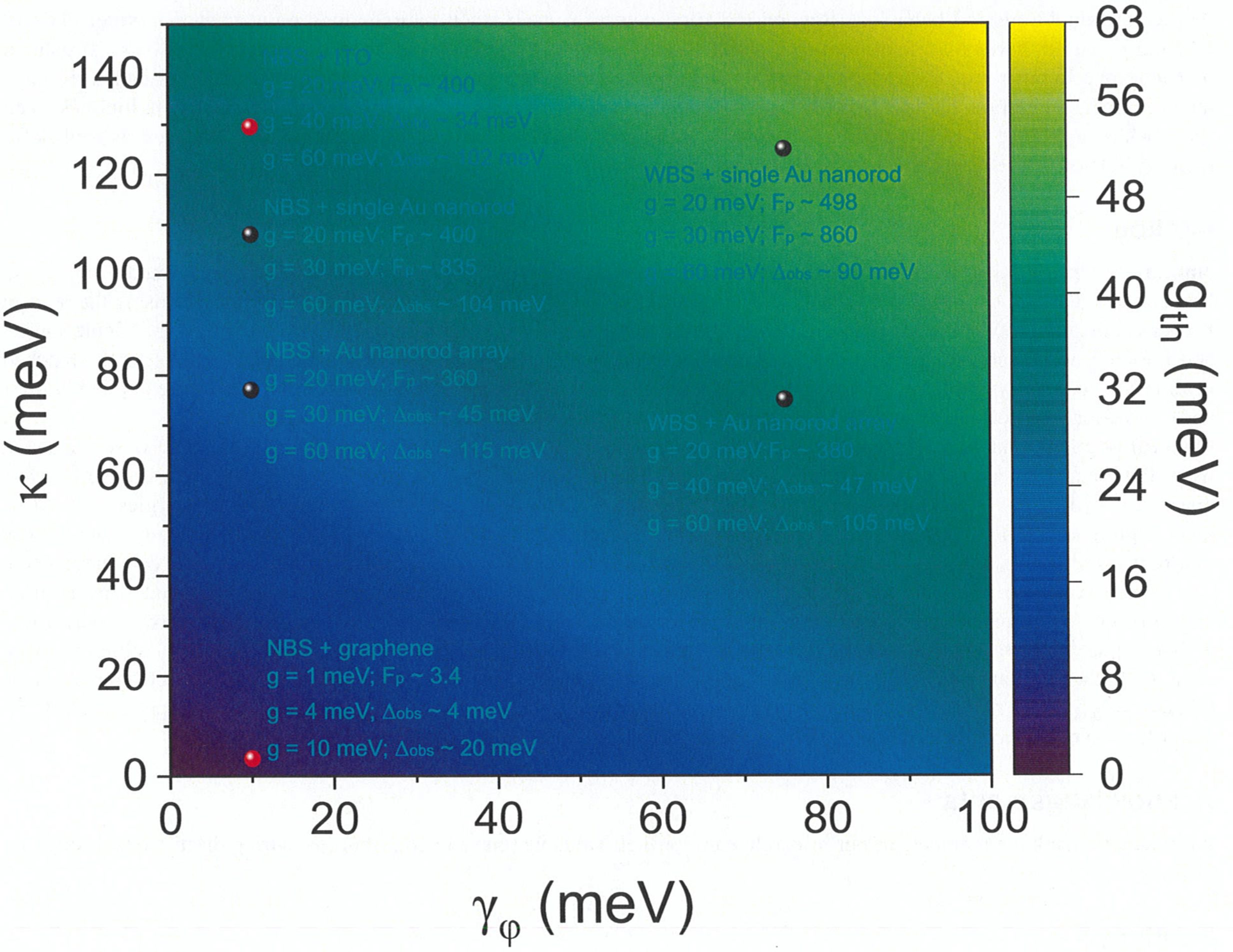

**Figure 5. Summary of the emitter-antenna parameter space explored in this work.** The diagram is plotted as a function of emitter pure dephasing, $\gamma_\phi$, and antenna linewidth, $\kappa$. The background color represents the threshold coupling strength, $g_{\text{th}}$, for the appearance of resolved strong-coupling signatures. Black points correspond to the Au-based reference antennas, whereas red points indicate the alternative conductive antennas based on ITO and graphene. The annotations list the coupling strengths $g$ tested in the simulations and mark whether we operate in weak-, intermediate-, or strong-coupling with resolved features.

## Discussion

This work provides a quantum-optical framework for understanding how scintillator NCs evolve from rate-engineered emission to hybrid light-matter dynamics under ionizing radiation. The results show that the decisive parameters are not only the interaction strength itself, but also the relative magnitudes of emitter dephasing and antenna linewidth, which determine whether coherent exchange becomes observable in practice. From this perspective, the most favorable regime is reached for narrow-band scintillators coupled to spectrally narrow antennas, with conductive platforms such as ITO and graphene offering particularly attractive opportunities in the near-infrared. More broadly, the present analysis identifies strong coupling as a realistic design target for scintillators and points toward nanoplasmonic strategies in which the radiation response is shaped not only by faster recombination but also by the formation of hybrid optical states. A particularly important outcome of this study is that the most favorable conditions for the emergence of strong-coupling signatures are not restricted to conventional noble-metal nanoantennas. While ITO demonstrates that conductive oxide antennas can already access this regime in the near-infrared ($g = 40$ meV), graphene provides the most extreme parameter set considered here, combining an ultranarrow antenna linewidth ($\kappa = 3.5$ meV) with the lowest threshold for observable coherent light-matter exchange ($g = 4$meV). More broadly, conductive nanoplasmonic platforms beyond ITO and graphene, such as titanium nitride, may provide an attractive route for future broadband architectures spanning the visible to the near-infrared, especially in combination with scintillators that exhibit dual-emission behavior. Beyond radiation detection itself, such regimes may point toward a wider class of ionizing-radiation-driven plasmonic concepts in which scintillation is treated not only as a light source but also as a structured optical signal. In this sense, controllable hybrid optical response may offer an alternative route to temporally extended emission under ionizing excitation, distinct from the trap-mediated mechanisms responsible for conventional persistent luminescence. Such behavior may be of future interest for radiative energy-conversion platforms related to nuclear batteries, as well as for memory-related or spectrally encoded radiation-imaging concepts.

## Methods

**Numerical Simulations.** Three-dimensional finite-difference time-domain (FDTD) simulations were performed using Ansys Lumerical. Literature data were used for the dielectric functions of Au[38] and ITO[39]. Graphene was modeled using the built-in Lumerical material model based on the Falkovsky conductivity formalism[40]. Excitation was introduced using a dipole source, and a nonuniform mesh with a 0.5 nm refinement region was applied around the nanoplasmonic structures. Perfectly matched layer (PML) boundary conditions were used for isolated structures, whereas Bloch boundary conditions were applied in the $x$ and $y$ directions for periodic structures supporting surface lattice resonances (SLRs).
**Quantum-optical modeling.** The first-order coherence function $g^{(1)}(\tau)$ and the corresponding detected spectra were calculated using Python and the QuTiP quantum-optics library[41,42]. The emitter-antenna system was described using a standard driven-dissipative two-level-system-antenna model, with all parameters, including the bare emitter and antenna energies, linewidths, and coupling strength $g$, taken directly from experimental fits and FDTD-derived mode properties. The steady-state density matrix was obtained using `qutip.steadystate`. The two-time correlation function $\langle a^\dagger(t)a(t+\tau)\rangle$ was calculated using `correlation_2op_1t`, and $g^{(1)}(\tau)$ was obtained by normalization. The detected spectra were evaluated by Fourier-transforming the corresponding correlation functions using the Wiener-Khinchin relation. All simulations were performed without adjustable free parameters, using the same detuning and linewidth values extracted from the literature radioluminescence datasets. In the calculated $S_{\mathrm{det}}$ maps, equal weighting factors were assigned to the antenna and direct-emitter detection channels ($F_{\mathrm{antenna}} = F_{\mathrm{rad}}$). This choice was made to isolate the influence of detuning and light-matter coupling on the spectral lineshape, rather than to reproduce the exact collection efficiencies of a particular radioluminescence experiment.

## Acknowledgements

We sincerely thank the members of our research team for their valuable time and contributions during discussions throughout this research.

## Funding

All authors acknowledge research funds from the National Science Centre, Poland, under grant OPUS-24 no. 2022/47/B/ST5/01966.

## Author contributions statement

**M.M.**: Methodology, Conceptualization, Quantum-optical modeling, Writing-original draft, Writing-review & editing, **D.K.**: Conceptualization, FDTD simulations, Writing-review and editing; **M.D.B.**: Conceptualization, Supervision, Resources, Writing-review & editing, Project administration, Funding acquisition.

# Supplementary Information for Tuning light-matter interaction of near-infrared nanoplasmonic scintillators

**Michał Makowski[1,2,*], Dominik Kowal[1], and Muhammad Danang Birowosuto[1]**

[1]Łukasiewicz Research Network - PORT Polish Center for Technology Development, Wrocław, 54-066, Poland
[2]Institute of Physics, Faculty of Physics, Astronomy and Informatics, Nicolaus Copernicus University, Grudziadzka 5, 87-100 Torun, Poland
[*]michal.makowski@port.lukasiewicz.gov.pl

## ABSTRACT

## List of Figures

## List of Tables

## Quantum-optical model and numerical implementation

The emitter-antenna system was modeled within the framework of open quantum systems using a driven-dissipative Jaynes-Cummings formalism[1–3]. The emitter was treated as an effective two-level system coupled to a single confined optical mode representing the plasmonic antenna. In the rotating frame used in the simulations, the Hamiltonian was written as

$$\hat{H} = \hbar\Delta_c \hat{a}^\dagger \hat{a} + \hbar g \left(\hat{a}^\dagger \hat{\sigma} + \hat{a}\hat{\sigma}^\dagger\right), \tag{S1}$$

where $\hat{a}$ and $\hat{a}^\dagger$ denote the annihilation and creation operators of the antenna mode, $\hat{\sigma}$ and $\hat{\sigma}^\dagger$ are the lowering and raising operators of the emitter, $\hat{\sigma}_z$ is the Pauli $z$ operator of the emitter, $g$ is the light-matter coupling strength, and $\Delta_c = \omega_c - \omega_e$ is the detuning between the antenna and emitter transition energies.

The system dynamics were described by the Lindblad master equation

$$\frac{d\hat{\rho}}{dt} = -\frac{i}{\hbar}[\hat{H},\hat{\rho}] + \sum_i \mathscr{L}_i[\hat{\rho}], \tag{S2}$$

where $\hat{\rho}$ is the density matrix and the Lindblad superoperators account for antenna loss, emitter decay, pure dephasing, and incoherent pumping. Each dissipative channel was introduced in the standard Lindblad form

$$\mathscr{L}_{\hat{C}}[\hat{\rho}] = \hat{C}\hat{\rho}\hat{C}^\dagger - \frac{1}{2}\left(\hat{C}^\dagger\hat{C}\hat{\rho} + \hat{\rho}\hat{C}^\dagger\hat{C}\right), \tag{S3}$$

where $\hat{C}$ denotes the corresponding collapse operator. In the numerical implementation, the collapse operators were taken as

$$\sqrt{\kappa}\hat{a}, \qquad \sqrt{\gamma}\hat{\sigma}, \qquad \sqrt{\gamma_\phi/4}\,\hat{\sigma}_z, \qquad \sqrt{P_c}\hat{a}^\dagger, \qquad \sqrt{P_x}\hat{\sigma}^\dagger, \tag{S4}$$

where $\kappa$ is the antenna loss rate, $\gamma$ is the emitter radiative decay rate, $\gamma_\phi$ is the pure dephasing rate, and $P_c$ and $P_x$ are the incoherent pumping rates of the antenna and emitter channels, respectively. In practice, only operators corresponding to non-zero rates were included in the simulations.

With this choice of collapse operators, the decay of emitter coherence is governed by the transverse relaxation rate

$$\Gamma_2 = \frac{\gamma}{2} + \gamma_\phi. \tag{S5}$$

Accordingly, the effective homogeneous linewidth of the emitter is determined not only by the radiative decay rate $\gamma$, but by the combined contribution of population decay and pure dephasing. In the parameter regimes considered here, the emitter linewidth was therefore controlled predominantly by $\gamma_\phi$ whenever $\gamma \ll \gamma_\phi$.

The steady-state density matrix $\hat{\rho}_{ss}$, defined by the condition

$$\frac{d\hat{\rho}_{ss}}{dt} = 0, \tag{S6}$$

was obtained numerically and used as the reference state for the evaluation of temporal and spectral observables. The temporal response was analyzed through the first-order correlation function of the detected field,

$$G^{(1)}(\tau) = \langle \hat{E}^\dagger_{det}(\tau)\hat{E}_{det}(0)\rangle_{ss}, \tag{S7}$$

where the subscript ss denotes averaging in the steady state. The detected field operator was defined as

$$\hat{E}_{det} = \sqrt{F_{antenna}\kappa}\,\hat{a} + \sqrt{F_{rad}\gamma}\,\hat{\sigma}, \tag{S8}$$

with $F_{antenna}$ and $F_{rad}$ representing the relative weights of the antenna-like and emitter-like detection channels. For visualization, the calculated temporal correlation was normalized to its zero-delay magnitude, and a global phase rotation was applied so that the plotted traces represent the real, imaginary, or absolute parts of the normalized first-order coherence function.

The spectral response was evaluated from steady-state two-time correlation functions using the standard quantum-optical relation between emission spectra and Fourier-transformed field correlations. In the implementation used here, the frequency variable $\omega$ was defined relative to the antenna resonance. The antenna-like contribution was calculated as

$$S_{antenna}(\omega) \propto F_{antenna}\frac{\kappa}{\pi}\,\mathrm{Re}\left[\int_0^\infty \langle \hat{a}^\dagger(\tau)\hat{a}(0)\rangle_{ss} e^{i\omega\tau}\,d\tau\right], \tag{S9}$$

while the emitter-like contribution was written as

$$S_{\mathrm{rad}}(\omega) \propto F_{\mathrm{rad}} \frac{\gamma}{\pi} \mathrm{Re}\left[ \int_0^{\infty} \langle \hat{\sigma}^\dagger(\tau) \hat{\sigma}(0) \rangle_{\mathrm{ss}} e^{i\omega\tau} \, d\tau \right]. \tag{S10}$$

The detected spectrum was then constructed as an incoherent sum of these two channels,

$$S_{\mathrm{det}}(\omega) = S_{\mathrm{antenna}}(\omega) + S_{\mathrm{rad}}(\omega). \tag{S11}$$

Accordingly, the temporal response and the spectral response were not evaluated from exactly the same composite detected-field expression: the first-order temporal correlation was calculated using the mixed detected-field operator $\hat{E}_{\mathrm{det}}$, whereas the detected spectrum was assembled as the sum of separately evaluated antenna-like and emitter-like spectral contributions. In this form, interference terms between the two detection channels were not included explicitly in the spectral decomposition.

All simulations were performed in Python using the QuTiP library[4,5]. The steady-state density matrix was obtained with `qutip.steadystate`, the first-order temporal correlation function was calculated using `correlation_2op_1t`, and the spectra were evaluated with `qutip.spectrum`. The antenna Hilbert space was truncated to a finite number of Fock states, and the numerical solver settings were chosen to ensure convergence of both the temporal correlation functions and the spectra. The temporal axis was expressed in physical time units through the energy-time relation set by $\hbar$.

## Scattering spectra of plasmonic nanoparticles

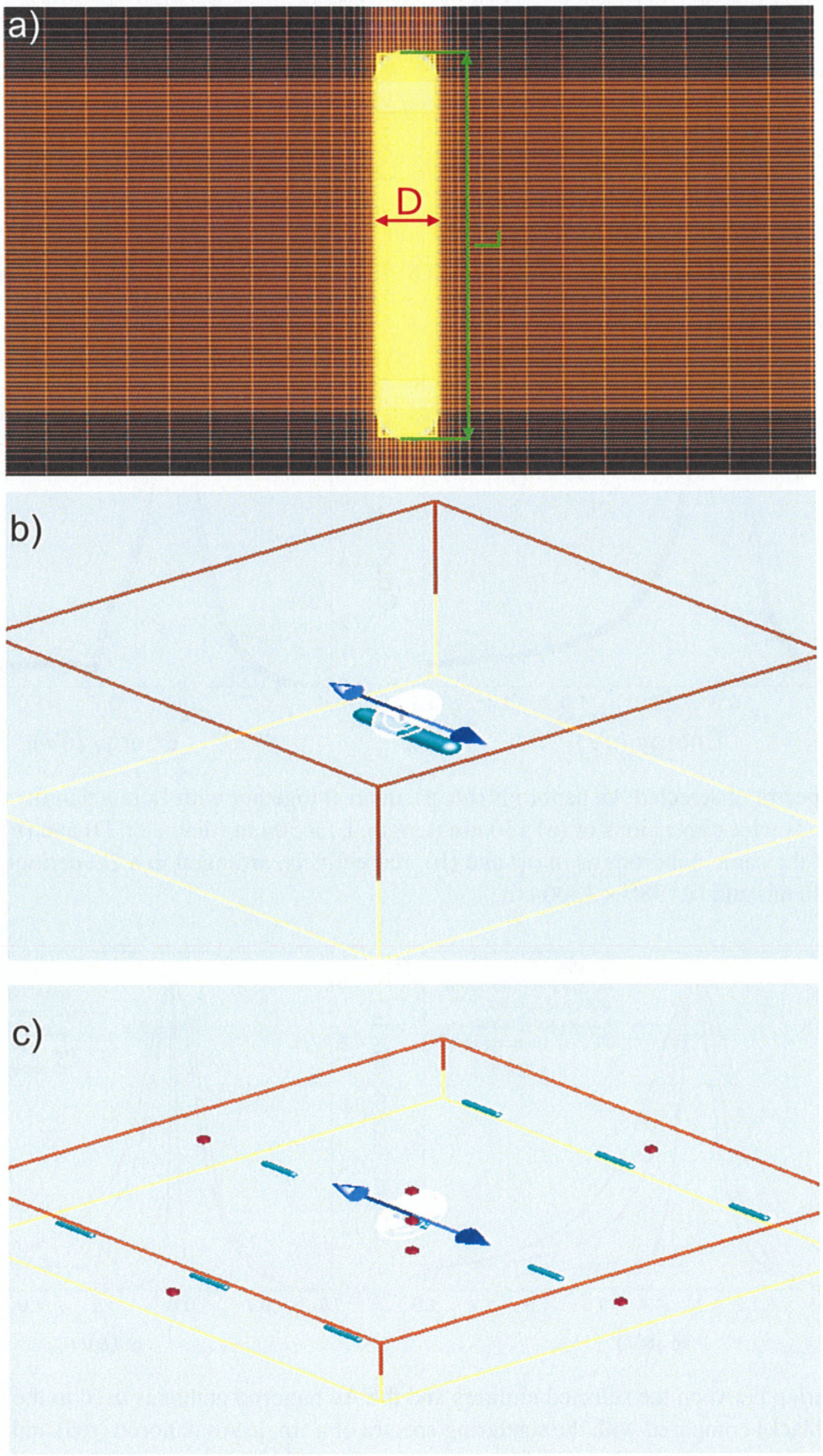

**Figure S1.** Schematic representation of the simulated Au nanorod structures and excitation geometry. **a**, Single Au nanorod of length $L$ and diameter $D$, overlaid with the simulation mesh (orange lines). **b**, Single Au nanorod illuminated by a plane-wave source (gray region). **c**, Periodic Au nanorod array defined by lattice periods $X$ and $Y$.

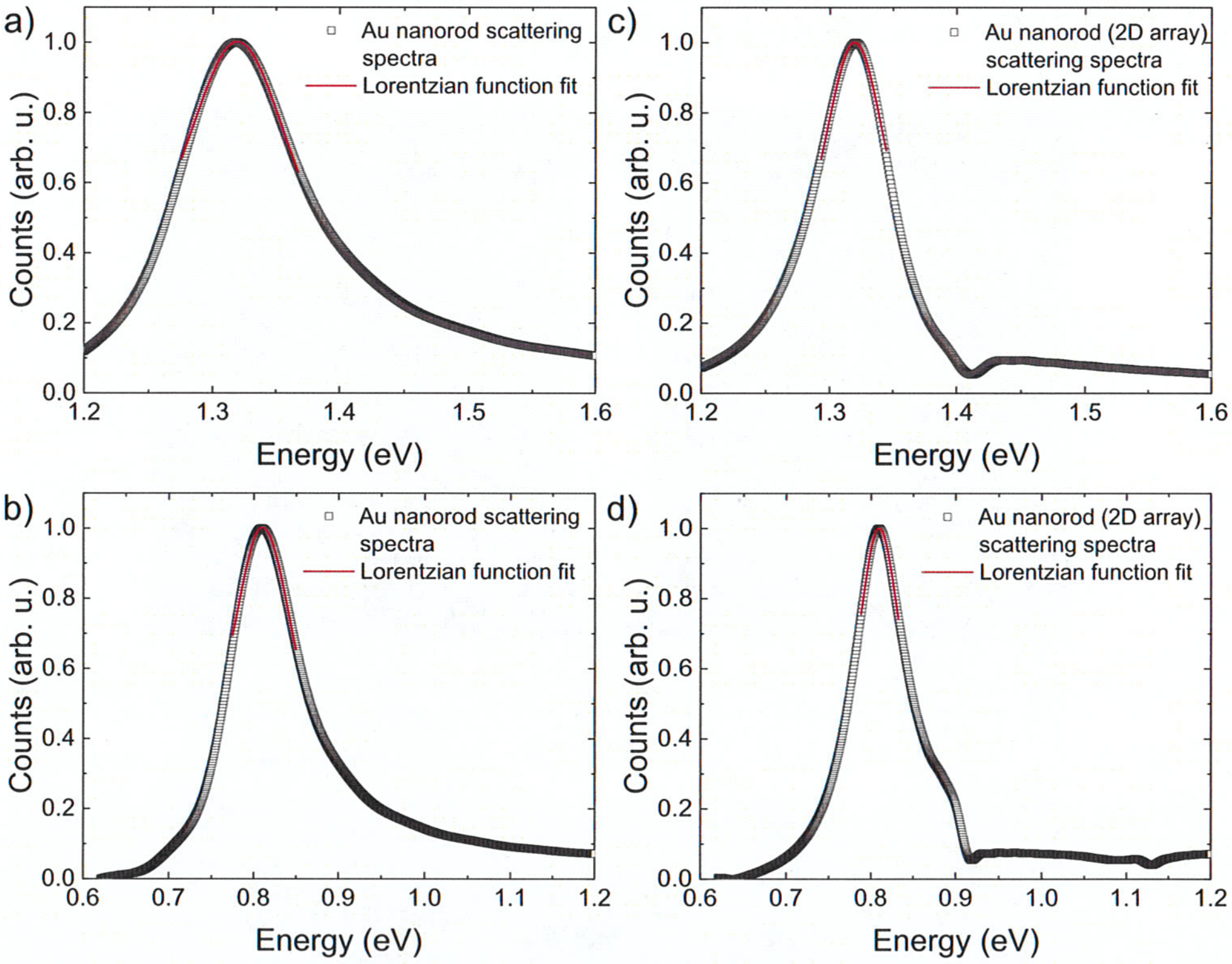

**Figure S2.** Scattering spectra of selected Au nanorods (black squares) together with Lorentzian fits (red lines). **a,b,** Scattering spectra of isolated nanorods with dimensions of (**a**) 130 nm (length, L) × 30 nm (diameter, D) and (**b**) 240 nm (L) × 30 nm (D). **c,d,** Scattering spectra of the same nanorods as in (**a**) and (**b**), respectively, arranged in a 2D periodic array with lattice constants of (**c**) 600 × 880 nm and (**d**) 980 × 1300 nm.

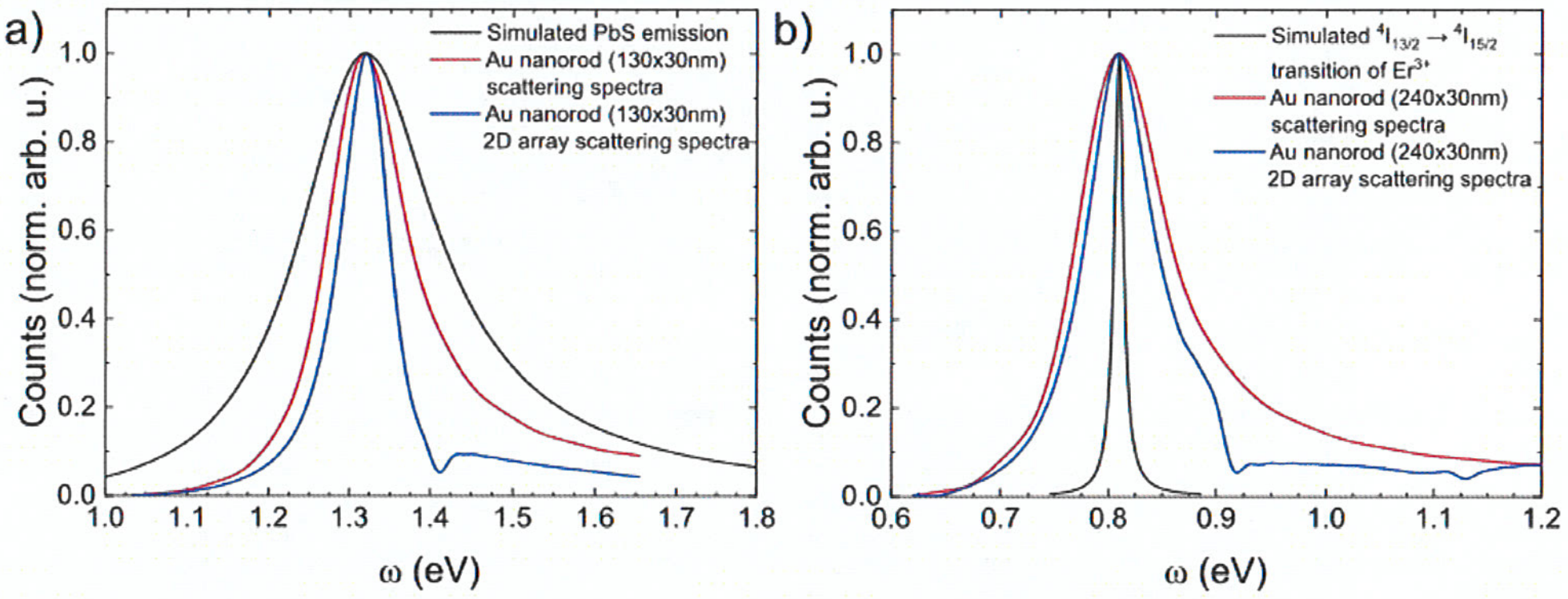

**Figure S3.** Spectral overlap between the selected emitters and the Au nanorod antennas used in the simulations. **a**, Simulated PbS emission spectrum (black) compared with the scattering spectra of a single Au nanorod (red) and a periodic 2D array of the same nanorods (blue). **b**, Simulated $Er^{3+}$ $^4I_{13/2} \rightarrow {}^4I_{15/2}$ emission band (black) compared with the scattering spectra of a single Au nanorod (red) and a periodic 2D array of the same nanorods (blue). The spectra illustrate the spectral matching conditions used to define broad- and narrow-band antenna for the wide-band and narrow-band emitter cases.

## Parameter extraction

The parameters used in the quantum-optical simulations were derived from the experimentally motivated emitter properties and from the simulated scattering spectra of the Au nanorod antennas. The antenna resonance energy $\omega_c$ was taken as the energy of the scattering maximum, while the antenna linewidth $\kappa$ was identified with the full width at half maximum (FWHM) of the corresponding plasmonic resonance,

$$\kappa = \mathrm{FWHM_c}. \tag{S12}$$

The emitter transition energy $\omega_e$ was taken from the maximum of the corresponding model emission spectrum, and the spectral detuning between emitter and antenna was defined as

$$\Delta = |\omega_c - \omega_e|. \tag{S13}$$

The emitter radiative decay rate $\gamma$ was estimated from the experimentally determined average emission decay time $\langle\tau\rangle$ using the energy-time relation

$$\gamma = \frac{\hbar}{\langle\tau\rangle}, \tag{S14}$$

with $\hbar = 0.658212$ meV ps and $\langle\tau\rangle$ expressed in ps. The pure dephasing rate $\gamma_\phi$ was then chosen so as to reproduce the experimental emitter linewidth within the present Lindblad formulation. For the collapse operators used in the model, the emitter coherence decays with the transverse relaxation rate

$$\Gamma = \frac{\gamma}{2} + \gamma_\phi. \tag{S15}$$

Accordingly, the homogeneous emitter linewidth is given by

$$\mathrm{FWHM_e} = 2\Gamma = \gamma + 2\gamma_\phi. \tag{S16}$$

Thus, for a given experimental emitter linewidth, the pure dephasing rate was obtained as

$$\gamma_\phi = \frac{\mathrm{FWHM_e} - \gamma}{2}. \tag{S17}$$

In the limit of long-lived emitters, where $\gamma \ll \mathrm{FWHM_{em}}$, this reduces to the approximate relation

$$\gamma_\phi \approx \frac{\mathrm{FWHM_e}}{2}. \tag{S18}$$

Therefore, in the present model, the antenna linewidth is introduced directly as $\kappa = \mathrm{FWHM_c}$, whereas the emitter linewidth is distributed between the radiative decay rate $\gamma$ and the pure dephasing rate $\gamma_\phi$ according to the relations above.

**Table S1.** Parameters used in the quantum-optical simulations for the selected emitter-antenna systems. WBS corresponds to PbS and NBS to $Lu_2O_3$:$Er^{3+}$. The definitions of all quantities and the parameter-extraction procedure are given in the. The uncertainties in the extracted plasmonic parameters are below 5%.

| Material | $\omega_c$ (eV) | $\kappa$ (meV) | $\omega_e$ (eV) | $\gamma_\phi$ (meV) | $\langle\tau\rangle$ ($\mu$s) | $\gamma$ (meV) | $\Delta$ (meV) |
|---|---|---|---|---|---|---|---|
| Au nanorod ($L = 130$ nm, $D = 30$ nm), single | 1.320 | 125 | 1.319 (WBS) | 75 | 4.2[6] | $1.57 \times 10^{-7}$ | 1 |
| Au nanorod ($L = 130$ nm, $D = 30$ nm), 2D array ($X = 600$ nm, $Y = 880$ nm) | 1.320 | 75.1 | 1.319 (WBS) | 75 | 4.2[6] | $1.57 \times 10^{-7}$ | 1 |
| Au nanorod ($L = 240$ nm, $D = 30$ nm), single | 0.810 | 108 | 0.809 (NBS) | 9.9 | 0.6[7] | $1.1 \times 10^{-9}$ | 1 |
| Au nanorod ($L = 240$ nm, $D = 30$ nm), 2D array ($X = 980$ nm, $Y = 1300$ nm) | 0.809 | 77.1 | 0.809 (NBS) | 9.9 | 0.6[7] | $1.1 \times 10^{-9}$ | 1 |

## Detuning-dependent evolution of the detected emission spectra

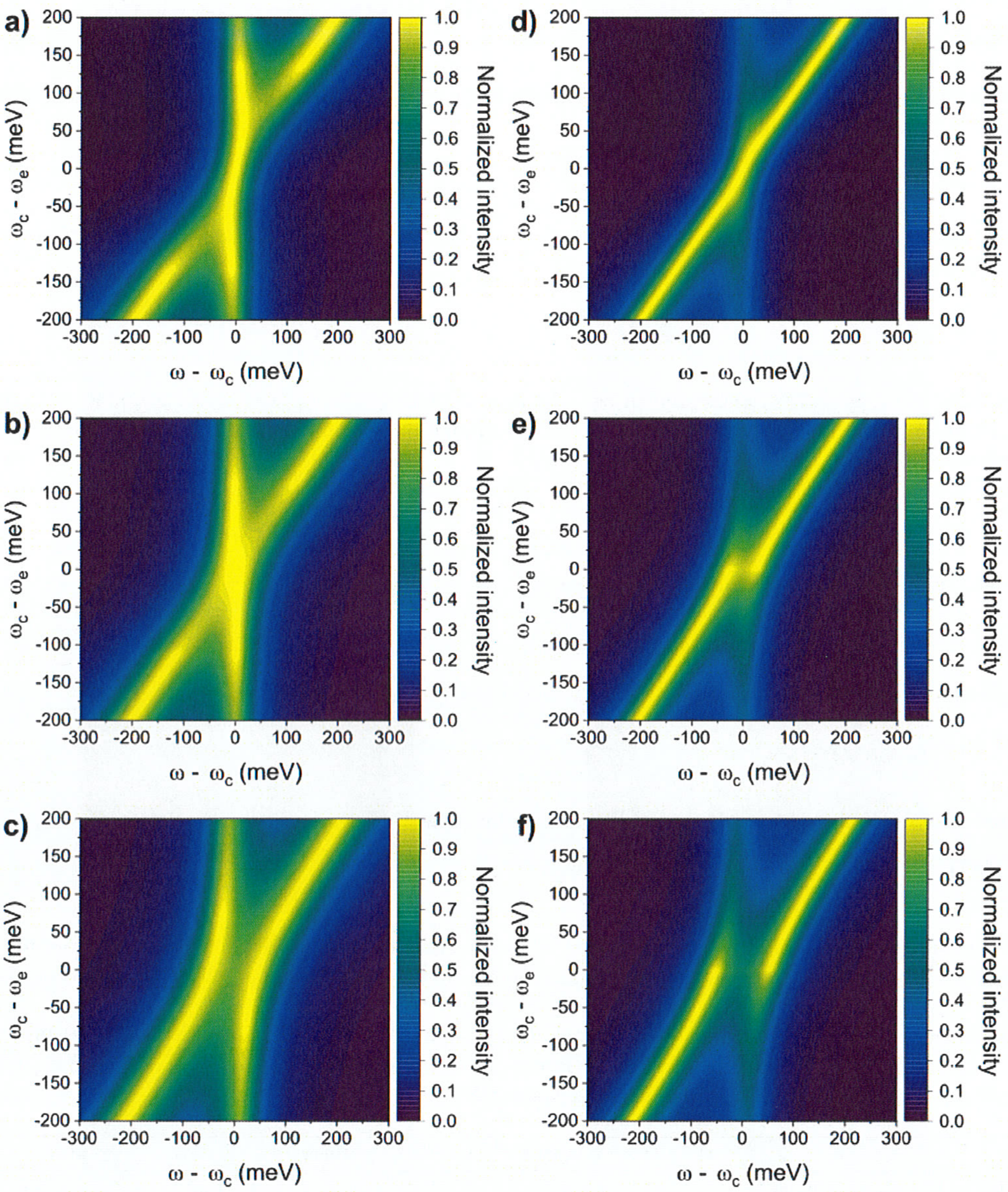

**Figure S4.** Normalized spectral maps of the detected emission $S_{\text{det}}$ for a wide-band scintillator (WBS) coupled to Au nanorods, plotted as a function of emission energy $(\omega - \omega_c)$ and detuning $(\omega_c - \omega_e)$. Panels (a,b,c) show the wider-band antenna, and panels (d,e,f) show the narrower-band antenna. The coupling strength is $g = 20$ meV in (a,d), $g = 40$ meV in (b,e), and $g = 60$ meV in (c,f). Each spectrum was normalized independently to highlight the evolution of the spectral shape. The onset of anticrossing becomes more evident with increasing $g$ and is more clearly resolved for the narrower-band antenna.

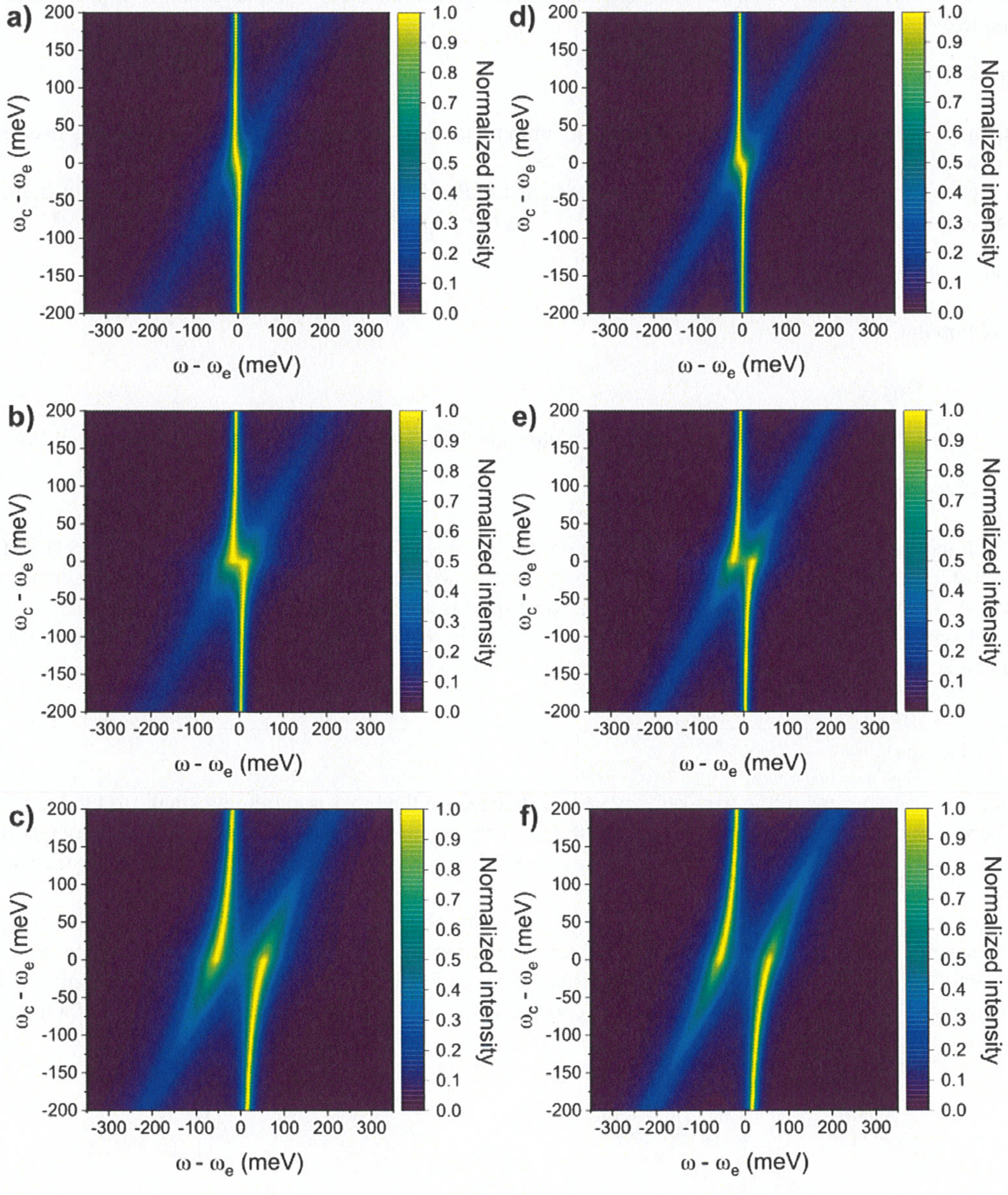

**Figure S5.** Normalized spectral maps of the detected emission $S_{\mathrm{det}}$ for a narrow-band scintillator (NBS) coupled to Au nanorods, plotted as a function of emission energy $(\omega - \omega_c)$ and detuning $(\omega_c - \omega_e)$. Panels (a,b,c) show the wider-band antenna, and panels (d,e,f) show the narrower-band antenna. The coupling strength is $g = 20$ meV in (a,d), $g = 40$ meV in (b,e), and $g = 60$ meV in (c,f). Each spectrum was normalized independently to highlight the evolution of the spectral shape. The onset of anticrossing becomes more evident with increasing $g$ and is more clearly resolved for the narrower-band antenna.

# 1 Calculation of temporal decay traces

The general quantum-optical model, parameter extraction, and numerical implementation are described above. Here, we summarize only the procedure used to calculate the transient decay traces and the corresponding effective Purcell-type enhancement factors.

To simulate the decay dynamics, the emitter-antenna system was initialized in a prepared state with the emitter excited and the cavity mode empty,

$$|\psi(0)\rangle = |e,0\rangle. \tag{S19}$$

In this transient branch of the calculation, the incoherent pumping terms were set to zero, so that only cavity loss, emitter decay, and pure dephasing were retained.

The subsequent time evolution was obtained from the same Lindblad master equation as used for the steady-state calculations. From the time-dependent density matrix, we evaluated the emitter population

$$n_{\mathrm{emit}}(t) = \langle \hat{\sigma}^{\dagger}\hat{\sigma}\rangle_t \tag{S20}$$

and the cavity population

$$n_{\mathrm{cav}}(t) = \langle \hat{a}^{\dagger}\hat{a}\rangle_t. \tag{S21}$$

To quantify the effective decay rate, the transient emitter population was fitted with a single-exponential function,

$$n_{\mathrm{emit}}(t) = A\exp[-\Gamma(g)t], \tag{S22}$$

where $A$ is the fitted amplitude and $\Gamma(g)$ is the effective decay rate obtained for a given coupling strength $g$. The fit was applied only after a short initial time cutoff and only over the range where the emitter population remained above a fixed fraction of its maximum value, in order to avoid numerical artifacts at very short times and low-signal tails at late times.

The Purcell-type enhancement values reported in this work were then obtained by normalizing the fitted decay rates to the value calculated at the lowest considered coupling strength, $g = 1$ meV,

$$F_{\mathrm{P}}(g) = \frac{\Gamma(g)}{\Gamma(1\ \mathrm{meV})}. \tag{S23}$$

This reference was used instead of the bare radiative rate $\gamma/\hbar$, because the latter is extremely small for the long-lived emitters considered here and would therefore lead to unrealistically large enhancement factors. Using $\Gamma(1\ \mathrm{meV})$ as the reference provides a more conservative, physically meaningful estimate of the transient decay acceleration within the same modeling framework.

Within the weak-coupling regime, $F_{\mathrm{P}}$ can be interpreted as an effective Purcell-type enhancement of the emitter decay dynamics. At larger coupling strengths, where hybrid light-matter states begin to form, and the dynamics may deviate from a strictly irreversible spontaneous-emission picture, it should be understood more generally as a fitted temporal acceleration factor. We define the intermediate-coupling region for $g$ values at which the decay behavior ceases to be exponential. Obtained $F_{\mathrm{P}}$ are summarized in Fig. S6

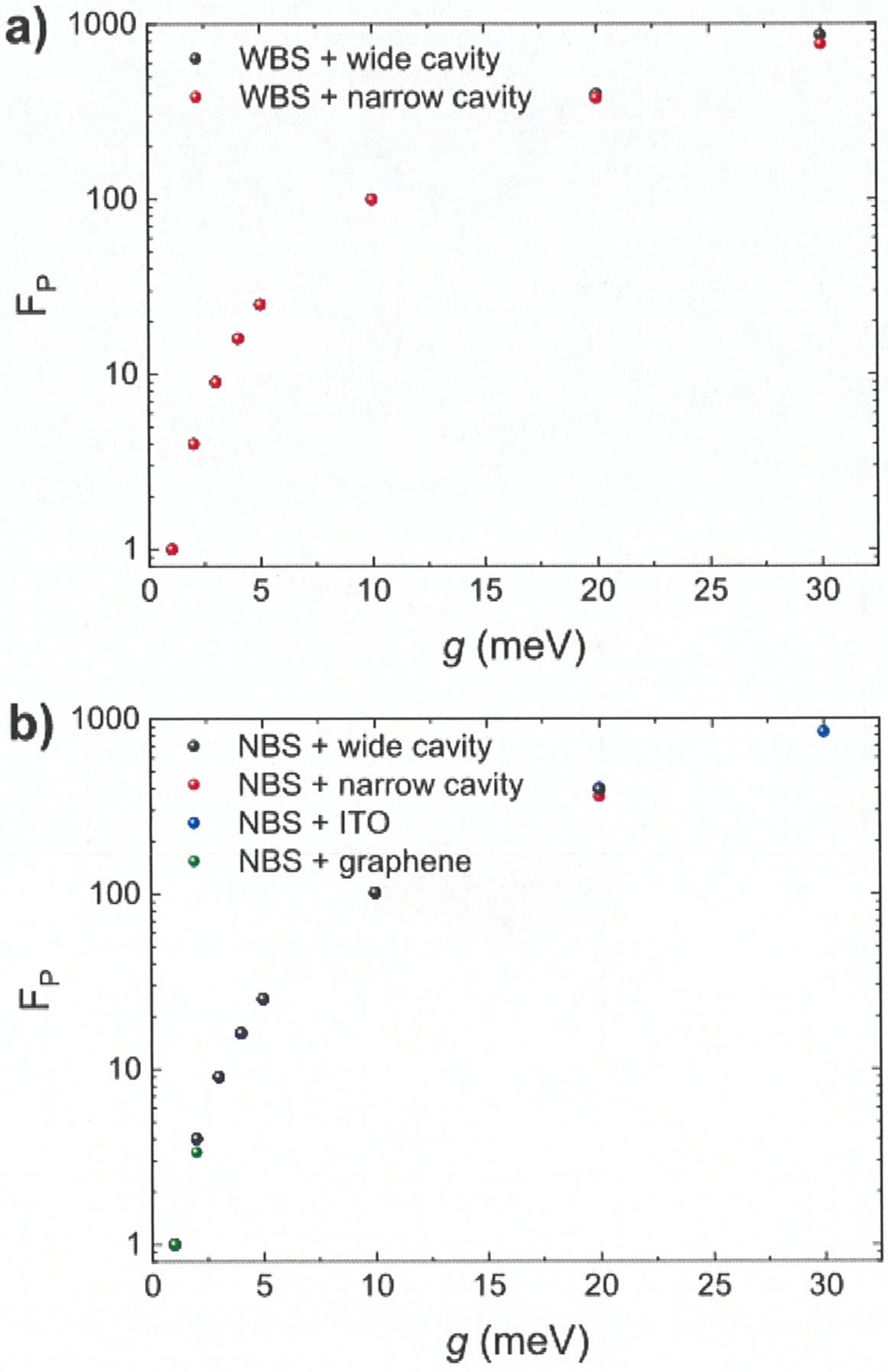

**Figure S6.** Calculated Purcell factors for WBS (a) and NBS (b).

## Alternative antennas

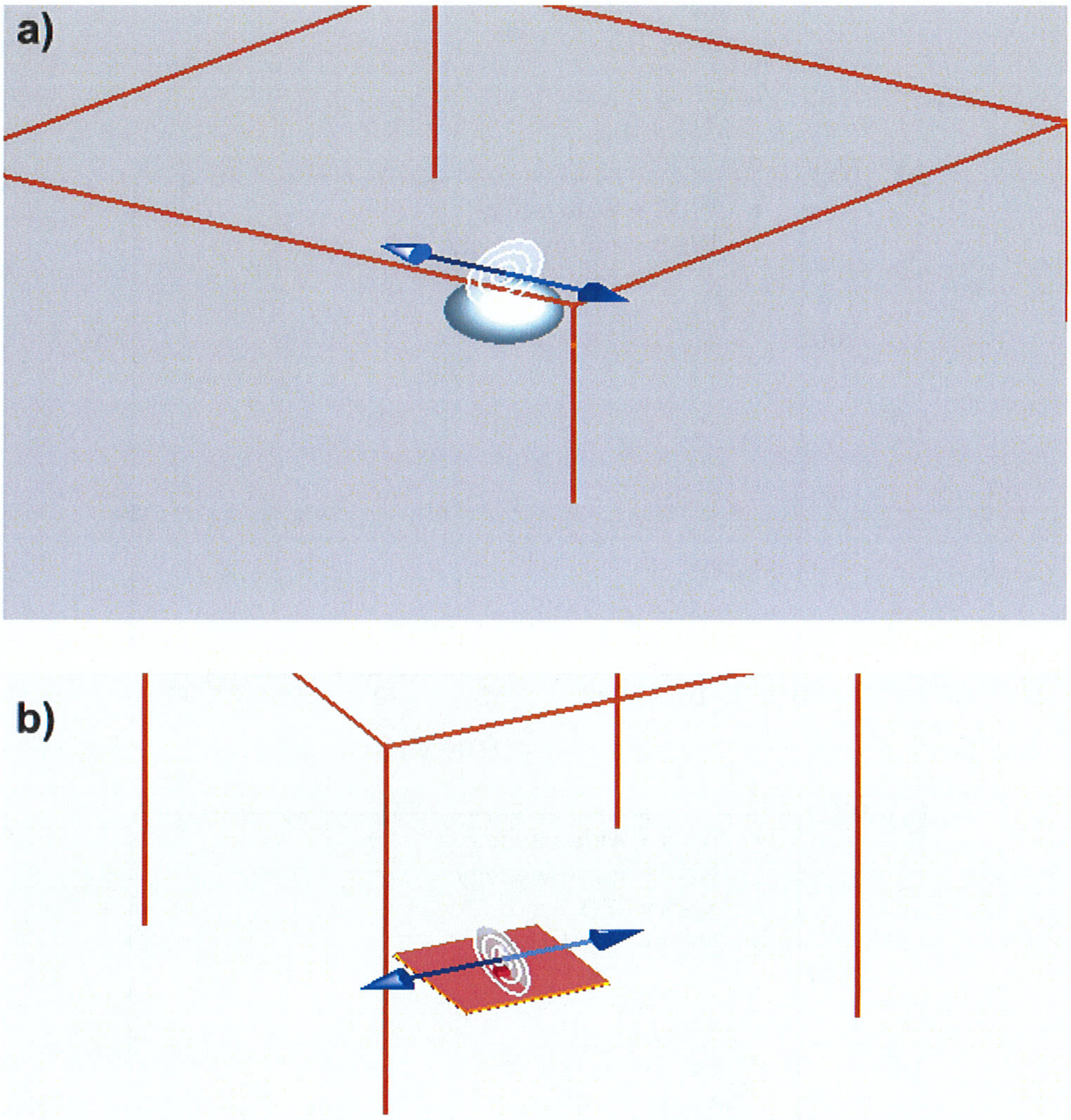

**Figure S7.** Schematic representation of the simulated ITO (a) and graphene (b) structures and excitation geometry

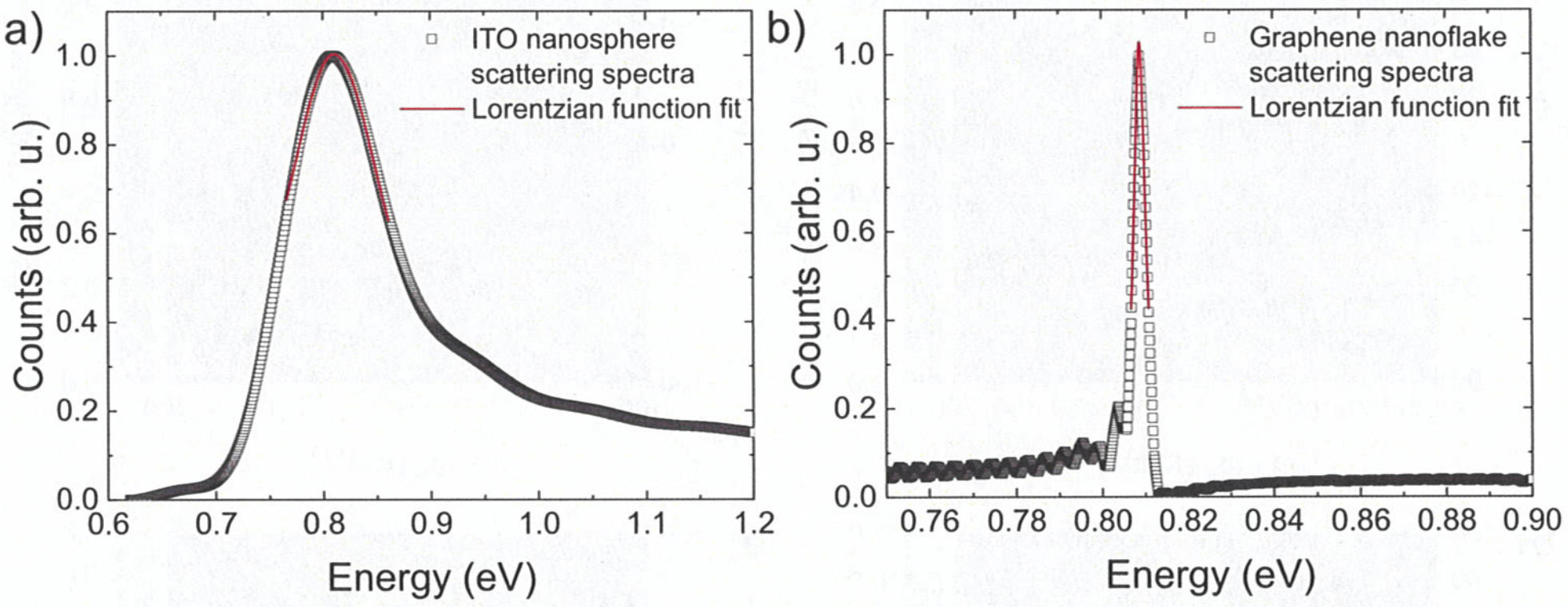

**Figure S8.** Scattering spectra of selected alternative antennas (black squares) together with Lorentzian fits (red lines). **a,** Scattering spectra of isolated ITO nanosphere ($r = 45$ nm). **b,** Scattering spectra of graphene nanoflake ($13 \times 13$ nm and thickness of 1 nm)

**Table S2.** Parameters used in the quantum-optical simulations for the selected alternative antennas. The definitions of all quantities and the parameter-extraction procedure are given in the. The uncertainties in the extracted plasmonic parameters are below 5%.

| Material | $\omega_c$ (eV) | $\kappa$ (meV) | $\omega_e$ (eV) | $\gamma_\phi$ (meV) | $\langle\tau\rangle$ ($\mu$s) | $\gamma$ (meV) | $\Delta$ (meV) |
|---|---|---|---|---|---|---|---|
| ITO nanosphere ($r = 45$ nm) | 0.810 | 129 | 0.809 (NBS) | 9.9 | 0.6[7] | $1.1 \times 10^{-9}$ | 1 |
| Graphene nanoflake ($13 \times 13$ nm and thickness of 1 nm) | 0.809 | 3.5 | 0.809 (NBS) | 9.9 | 0.6[7] | $1.1 \times 10^{-9}$ | 0 |

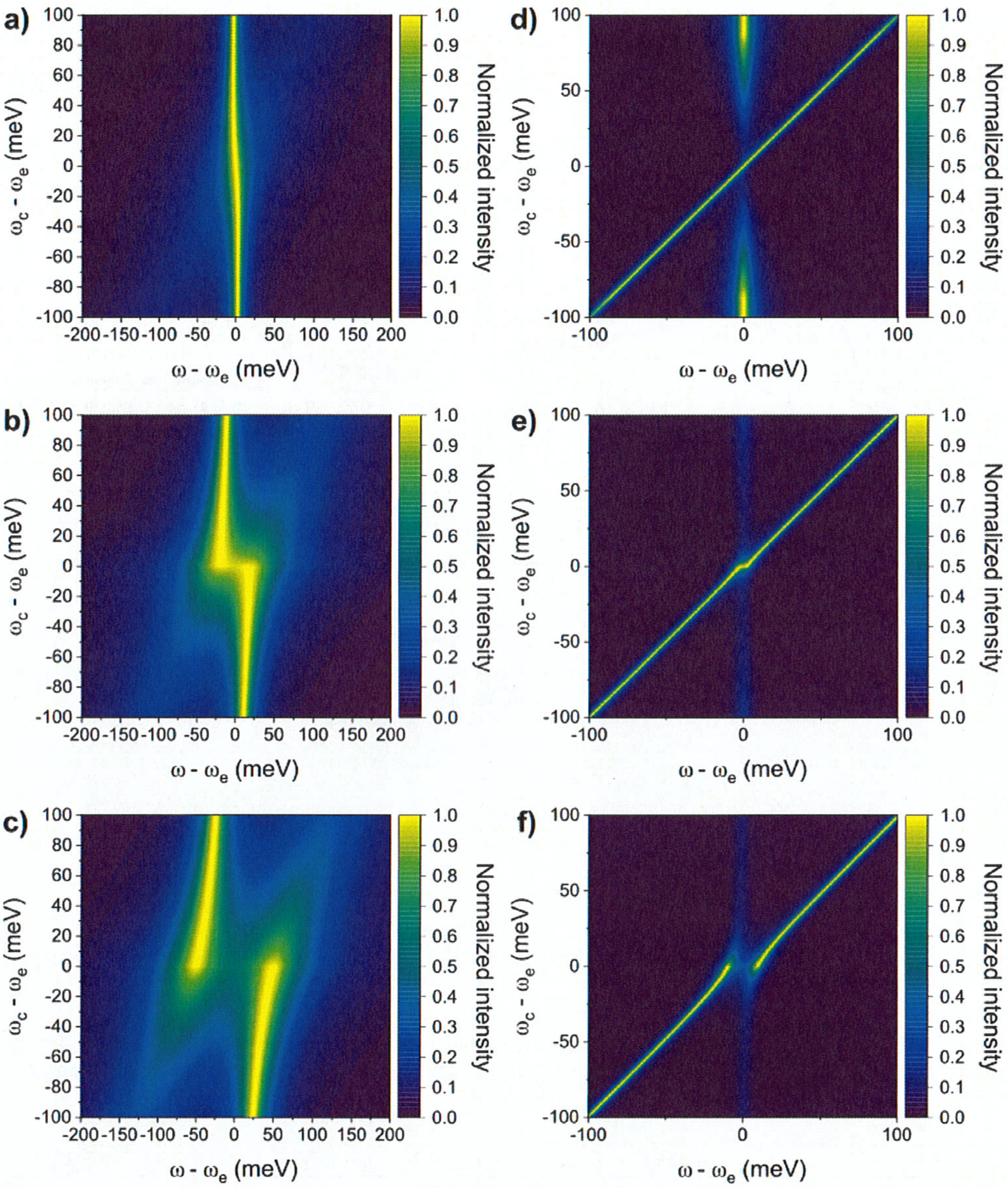

**Figure S9.** Normalized spectral maps of the detected emission $S_{\text{det}}$ for a narrow-band scintillator (NBS) coupled to alternative antennas, plotted as a function of emission energy $(\omega - \omega_c)$ and detuning $(\omega_c - \omega_e)$. Panels (a,b,c) show the ITO antenna, and panels (d,e,f) show the graphene antenna. The coupling strength is $g = 20$ meV in (a), $g = 40$ meV in (b), and $g = 60$ meV in (c) while $g = 1$ meV in (d), $g = 4$ meV in (e), and $g = 10$ meV in (f). Each spectrum was normalized independently to highlight the evolution of the spectral shape.